\documentstyle[12pt]{article} 
\begin{document}
\def\theequation{\arabic{section}.\arabic{equation}}
\newcommand{\be}{\begin{equation}} 
\newcommand{\ee}{\end{equation}}
\begin{titlepage} 
\setcounter{page}{1} 
\title{Einstein frame or Jordan frame~?} 
\author{Valerio Faraoni$^{1}$ and Edgard Gunzig$^{1,2}$\\ \\
{\small \it $^1$ Research Group in General Relativity}\\
{\small \it Universit\'{e} Libre de Bruxelles, Campus Plaine CP~231}\\ 
{\small \it Blvd. du Triomphe, 1050 Bruxelles, Belgium}\\ 
{\small \it $^2$ Instituts Internationaux de Chimie et de Physique Solvay}
} \date{} 
\maketitle
\vspace*{1truecm} 
\begin{abstract} Scalar--tensor
theories of gravity can be formulated in the Jordan or in the Einstein
frame, which are conformally related. The issue of which conformal frame
is physical is a contentious one; we provide a straightforward example
based on gravitational waves in order to clarify the issue. 
\end{abstract} 
\vspace*{1truecm} \begin{center} 
Published in {\em Int. J. Theor. Phys.} {\bf 38}, 217.
\end{center}
\end{titlepage} \clearpage

\section{Introduction} 

Scalar--tensor theories of gravity, of which Brans--Dicke \cite{BD61}
theory is the prototype, are competitors to Einstein's theory of general
relativity for the description of classical gravity. Renewed interest in
scalar--tensor theories of gravity is motivated by the extended
\cite{extended} and hyperxtended \cite{hyperextended} inflationary
scenarios of the early universe. Additional motivation arises from the
presence of a fundamental Brans--Dicke--like field in most high
energy
physics theories unifying gravity with the other interactions (the string
dilaton, the supergravity partner of spin $1/2$ particles, etc.).

It is well known since the original Brans--Dicke paper \cite{BD61} that
two formulations of scalar--tensor theories are possible; the version in
the so--called Jordan conformal frame commonly presented in the textbooks
(e.g. \cite{Weinberg72}--\cite{Will93}), and the less known version based
on the Einstein conformal frame, which is related to the former one by a
conformal transformation and a redefinition of the gravitational scalar
field present in the theory. The possibility of two formulations related
by a conformal transformation exists also for Kaluza--Klein theories and
higher derivative theories of gravity (see
\cite{MagnanoSokolowski94,FGN98} for reviews). The problem of whether the
two formulations of a scalar--tensor theory in the two conformal frames
are equivalent or not has been the issue of lively debates, which are not
yet settled, and often is the source of confusion in the technical
literature.  While many authors support the point of view that the two
conformal frames are equivalent, or even that physics at the energy scale
of classical gravity and classical matter is always conformally invariant,
other
authors support the opposite point of view, and others again are not aware
of the problem (see the ``classification of authors'' in
\cite{MagnanoSokolowski94}). The issue is important in principle and in
practice, since there are many applications of scalar--tensor theories and
of conformal transformation techniques to the physics of the early
universe and to astrophysics. The theoretical predictions to be compared
with the observations (in cosmology, the existence of inflationary
solutions of the field equations, and the spectral index of density
perturbations)  crucially depend on the
conformal frame adopted to perform the calculations.

In addition, if the two formulations of a scalar--tensor theory are not
equivalent, the problem arises of whether one of the two is physically
preferred, and which one has to be compared with experiments and
astronomical observations. Are both conformal versions of the same theory
viable, and good candidates for the description of classical gravity~?
Unfortunately many authors neglect these problems, and the issue is not
discussed in the textbooks explaining scalar--tensor theories.  On the
other hand, it emerges from the work of several authors, in different
contexts (starting with Refs. \cite{KK,Cho94,Cho97} on Kaluza--Klein and
Brans--Dicke theories, and summarized in
\cite{MagnanoSokolowski94,FGN98}), that 

\begin{enumerate}

\item the formulations of a scalar--tensor theory in the two conformal
frames are physically inequivalent.

\item The Jordan frame formulation of a scalar--tensor theory is not
viable because the energy density of the gravitational scalar field
present in the theory is not bounded from below (violation of the weak
energy condition \cite{Wald84}). The system therefore is unstable and
decays toward a lower and lower energy state {\em ad infinitum}
\cite{MagnanoSokolowski94,FGN98}). 

\item The Einstein frame formulation of scalar--tensor theories is free
from the problem 2). However, in the Einstein frame there is a violation
of the equivalence principle due to the anomalous coupling of the scalar
field to ordinary matter (this violation is small and compatible with the
available tests of the equivalence principle \cite{Cho97}; it is indeed
regarded as an important low energy manifestation of compactified theories
\cite{Cho92}--\cite{Cho97}, \cite{EPviolation}). 

\end{enumerate}

It is clear that property 2) is not acceptable for a viable {\em
classical} theory of gravity (a quantum system, on the contrary, may have
states with negative energy density \cite{Witten82,FordRoman929395}). A
classical theory must have a ground state that is stable against small
perturbations.

In spite of this compelling argument, there is a tendency to ignore the
problem, which results in a uninterrupted flow of papers performing
computations in the Jordan frame. The use of the latter is also
implicitely supported by most textbooks on gravitational theories. Perhaps
this is due to reluctance in accepting a violation of the equivalence
principle, on philosophical and aesthetic grounds, or perhaps it is due
to the fact that the best discussions of this subject are rather
mathematical than physical in character, and not well known. In this
paper, we present a straightforward argument in favor of the
Einstein frame, in the hope to help settling the issue. 

In Sec.~2 we recall the relevant formulas. In Sec.~3 we present a simple
argument based on scalar--tensor gravitational waves, and section~4
contains a discussion and the conclusions. 

\section{Conformal frames}

The textbook formulation of scalar--tensor theories of gravity is the one
in the Jordan conformal frame, in which the action takes the
form\footnote{The metric signature is --~+~+~+, the Riemann tensor is
given in terms of the Christoffel symbols by ${R_{\mu\nu\rho}}^{\sigma}=
\Gamma^{\sigma}_{\mu\rho ,\nu}-\Gamma^{\sigma}_{\nu\rho ,\mu}+
\Gamma^{\alpha}_{\mu\rho}\Gamma^{\sigma}_{\alpha\nu}-
\Gamma^{\alpha}_{\nu\rho}\Gamma^{\sigma}_{\alpha\mu} $, the Ricci tensor
is $R_{\mu\rho}\equiv {R_{\mu\nu\rho}}^{\nu}$, and
$R=g^{\alpha\beta}R_{\alpha\beta}$. $\nabla_{\mu}$ is the covariant
derivative operator, $\Box \equiv g^{\mu\nu}\nabla_{\mu}\nabla_{\nu}$,
$\eta_{\mu\nu}=$diag$(-1,1,1,1)$, and we use units in which the speed of
light and Newton's constant assume the value unity.} \be \label{1}
S=\frac{1}{16\pi}\int d^4x \sqrt{-g} \left[ f( \phi) R -\frac{\omega( \phi
)}{\phi} g^{\alpha\beta} \nabla_{\alpha}\phi \nabla_{\beta}\phi +\Lambda (
\phi ) \right] + \int d^4x \sqrt{-g} \, {\cal L}_{matter} \; , \ee where
${\cal L}_{matter}$ is the Lagrangian density of ordinary matter, and the
couplings $f( \phi )$, $\omega ( \phi)$ are regular functions of the
scalar field $\phi$. Although our discussion applies to the generalized
theories described by the action (\ref{1}), for simplicity, we will
restrict ourselves to Brans--Dicke theory, in which $\omega $ and
$\Lambda$ are constants and we will omit the non--gravitational part of
the action, which is irrelevant for our purposes. We further assume that
$\Lambda =0$; the field equations then reduce to \be \label{JFFE1}
R_{\mu\nu}-\frac{1}{2} g_{\mu\nu} R= \frac{\omega}{\phi^2} \left(
\nabla_{\mu}\phi \nabla_{\nu} \phi -\frac{1}{2} g_{\mu\nu}
\nabla^{\alpha}\phi \nabla_{\alpha}\phi \right) +\frac{1}{\phi} \left(
\nabla_{\mu}\nabla_{\nu} \phi-g_{\mu\nu} \Box \phi \right)  \; , \ee \be
\label{JFFE2} \Box \phi +\frac{\phi R}{2\omega}=0 \; .  \ee It is well
known since the original Brans--Dicke paper \cite{BD61} that another
formulation of the theory is possible: the conformal transformation \be
\label{CT} g_{\mu\nu} \longrightarrow \tilde{g}_{\mu\nu}=\phi \,
g_{\mu\nu} \; , \ee and the scalar field redefinition \be
\label{SFredefinition} \phi \longrightarrow \tilde{\phi}=\int \frac{
\left( 2\omega+3 \right)^{1/2}}{\phi} \, d\phi \; , \ee (where $\omega
>-3/2$), recast the theory in the so--called Einstein conformal
frame\footnote{Also called ``Pauli frame'' in
Refs.~\cite{Cho92}--\cite{Cho97})}, in which the gravitational part of the
action becomes that of Einstein gravity plus a non self--interacting
scalar field\footnote{If the Lagrangian density ${\cal L}_{matter}$ of
ordinary matter is included in the original action, it will appear
multiplied by a factor $\exp \left( -\alpha \tilde{\phi} \right)$ in the
action (\ref{Einsteinaction}); this anomalous coupling is responsible for
a violation of the equivalence principle in the Einstein frame
\cite{Cho92}--\cite{Cho97}, \cite{EPviolation}).}, 
\be \label{Einsteinaction}
S=\int d^4x
\sqrt{-\tilde{g}} \left[ \frac{\tilde{R}}{16\pi} -\frac{1}{2} \,
\tilde{g}^{\mu\nu} \tilde{\nabla}_{\mu}\tilde{\phi}
\tilde{\nabla}_{\nu}\tilde{\phi} \right] \; . 
\ee 
The field equations are
the usual Einstein equations with the scalar field as a source, 
\be
\tilde{R}_{\mu\nu}-\frac{1}{2} \tilde{g}_{\mu\nu} \tilde{R}= 8\pi \left(
\tilde{\nabla}_{\mu}\tilde{\phi} \tilde{\nabla}_{\nu} \tilde{\phi}
-\frac{1}{2} \, \tilde{g}_{\mu\nu} \tilde{\nabla}^{\alpha} \tilde{\phi}
\tilde{\nabla}_{\alpha}\tilde{\phi} \right) \; , 
\ee 
\be \tilde{\Box}
\tilde{\phi}=0 \; .  
\ee 
It has been pointed out (see \cite{MagnanoSokolowski94,FGN98} for reviews)
that the Jordan frame
formulation of Brans--Dicke theory is not viable because the sign of
the kinetic term for the scalar field is not positive definite, and hence
the theory does not have a stable ground state. The system decays toward
lower and lower energy states without a lower bound. On the contrary, the
Einstein frame version of the theory possesses the desired stability
property. These features were first discovered in Kaluza--Klein and
Brans--Dicke theory \cite{KK,Cho94,Cho97}, and later
(\cite{MagnanoSokolowski94,FGN98} and references therein) in
scalar--tensor and
non--linear theories of gravity with Lagrangian density of the form ${\cal
L}=f( \phi, R)$. Despite this difficulty with the energy, the textbooks
still present the Jordan frame version of the theory without mention of
its Einstein frame counterpart. The technical literature is also haunted
by confusion on this topic, expecially in cosmological applications
\cite{MagnanoSokolowski94,FGN98}. Many authors perform calculations in
both conformal frames, while others support the use of the Jordan frame,
or even claim that the two frames are physically equivalent. The issue of
the conformal frame may appear a purely technical one, but it is indeed
very important, in principle, and because the physical predictions of a
classical theory of gravity, or of an inflationary cosmological scenario,
are deeply affected by the choice of the conformal frame. 
Here, we study the violation of the weak energy condition by {\em
classical} gravitational waves. It appears very hard to argue with
the energy
argument that leads to the choice of the Einstein frame
\cite{MagnanoSokolowski94}; moreover, the entire realm of
classical\footnote{Quantum states can violate the weak energy condition
\cite{Witten82,FordRoman929395}; in this paper, we restrict to classical
gravitational theories.} physics is not conformally invariant. The
literature on the topic is rather mathematical and abstract, and can be
easily missed by the physically--minded reader. In the next
section we propose a physical illustration of how the weak energy
condition is violated in the Jordan frame, but not in the Einstein frame.

\section{Gravitational waves in the Jordan and in the Einstein frame}
 
We begin by considering gravitational waves in the Jordan frame version of
Brans--Dicke theory. In a locally freely falling frame, the metric and the
scalar field are decomposed as follows \setcounter{equation}{0} \be
\label{JFmetric} g_{\mu\nu}=\eta_{\mu\nu}+h_{\mu\nu} \; , \ee \be
\label{JFscalar} \phi=\phi_0 +\varphi \; , \ee where $\eta_{\mu\nu} $ is
the Minkowski metric, $\phi_0$ is constant, and the wave--like
perturbations $h_{\mu\nu}$, $\varphi/\phi_0 $ have the same order of
magnitude, \be {\mbox O} \left( \frac{\varphi}{\phi_0} \right) =\mbox{O}
\left( h_{\mu\nu} \right)= \mbox{O} \left( \epsilon \right)  \ee in terms
of a smallness parameter $\epsilon$. The linearized field equations in the
Jordan frame \be R_{\mu\nu}=\frac{\partial_{\mu}\partial_{\nu}
\varphi}{\phi_0} \; , \ee \be \Box \varphi =0 \; , \ee allow the expansion
of $\varphi$ in monochromatic plane waves:  \be \label{planewaves}
\varphi=\varphi_0 \cos \left( k_{\alpha} x^{\alpha} \right) \; , \ee where
$\varphi_0$ is constant and $\eta_{\mu\nu} k^{\mu} k^{\nu}=0$. Now note
that, for any timelike vector $\xi^{\mu}$, the quantity
$T_{\mu\nu}\xi^{\mu} \xi^{\nu}$ (which represents the energy density of
the waves as seen by an observer with four--velocity $\xi^{\mu}$) is
given, to the lowest order, by \be \label{JFenegydensity}
T_{\mu\nu}\xi^{\mu} \xi^{\nu}=- \left( k_{\mu} \xi^{\mu} \right)^2
\frac{\varphi}{\phi_0} \; .  \ee This quantity oscillates, changing sign
with the frequency of $\varphi$ and therefore violating the weak energy
condition \cite{Wald84}. In addition, the energy density is not quadratic
in the first derivatives of the field, and this implies that the energy
density of the scalar field $\varphi$ is of order O($\epsilon$), while the
contribution of the tensor modes $h_{\mu\nu} $ is only of order
O($\epsilon^2$) (and is given by the Isaacson effective stress--energy
tensor \cite{MTW} $T_{\mu\nu}^{(eff)}[ h_{\alpha\beta}]$). The Jordan
frame formulation of Brans--Dicke theory somehow discriminates between
scalar and tensor modes. From an experimental point of view, this fact has
important consequences for the amplification induced by scalar--tensor
gravitational waves on the light propagating through them, and for ongoing
{\em VLBI} observations \cite{Faraoni96}--\cite{BraccoTeyssandier}. If the
Jordan frame formulation of scalar--tensor theories was the physical one,
astronomical observations could potentially detect the time--dependent
amplification induced by gravitational waves in a light beam, which is of
order $\epsilon$ \cite{Faraoni96}. If instead the Einstein frame
formulation of scalar--tensor theories is physical (which is the case,
as as we shall see in the
following), then the amplification effect is of order $\epsilon^2$, and
therefore undetectable \cite{Faraoni96}--\cite{BraccoTeyssandier}.

We now turn our attention to gravitational waves in the Einstein frame
version of Brans--Dicke theory. The metric and scalar field decompositions
\be \label{EFM} \tilde{g}_{\mu\nu}=\eta_{\mu\nu}+\tilde{h}_{\mu\nu} \; ,
\ee \be \label{EFSF} \tilde{\phi}=\tilde{\phi}_0+\tilde{\varphi} \; , \ee
where $\tilde{\phi}_0$ is constant and O$ ( \tilde{h}_{\mu\nu})=$O$(
\tilde{\varphi}/\tilde{\phi}_0 )= $O$( \epsilon )$, lead to the equations
\be \label{EFFE1} \tilde{R}_{\mu\nu}-\frac{1}{2}\, \tilde{g}_{\mu\nu}
\tilde{R}=8\pi \left( \tilde{T}_{\mu\nu} [ \tilde{\varphi} ]
+\tilde{T}_{\mu\nu}^{(eff)} [ \tilde{h}_{\mu\nu} ] \right) \; , \ee \be
\label{EFFE2} \tilde{\Box} \tilde{\varphi}=0 \; . \ee Here $
\tilde{T}_{\mu\nu} [ \tilde{\varphi} ] = \partial_{\mu} \tilde{\varphi}
\partial_{\nu} \tilde{\varphi}- \eta_{\mu\nu} \partial^{\alpha}
\tilde{\varphi} \partial_{\alpha}\tilde{\varphi}/2 $. Again, we consider
plane monochromatic waves \be \tilde{\varphi}=\tilde{\varphi}_0 \cos
\left( l_{\alpha} x^{\alpha} \right) \; , \ee where $\tilde{\varphi}_0$ is
a constant and $\eta_{\mu\nu} l^{\mu}l^{\nu}=0$. The energy density
measured by an observer with timelike four--velocity $\xi^{\mu}$ in the
Einstein frame is \be \tilde{T}_{\mu\nu}\xi^{\mu} \xi^{\nu}=\left[ l_{\mu}
\xi^{\mu} \tilde{\varphi}_0 \sin \left( l_{\alpha} x^{\alpha} \right)
\right]^2 +\tilde{T}_{\mu\nu}^{(eff)} [ \tilde{h}_{\alpha\beta} ] \,
\xi^{\mu} \xi^{\nu} \; , \ee which is positive definite. The contributions
of the scalar and tensor modes to the total energy density have the same
order of magnitude O$( \epsilon^2 )$, and are both quadratic in the first
derivatives of the fields. The weak energy condition is satisfied in the
Einstein, but not in the Jordan frame; physically reasonable matter in the
classical domain is expected to satisfy the energy conditions
\cite{Wald84}.

Let us return for a moment to the Jordan frame: analogously to
eq.~(\ref{JFenegydensity}), the energy--momentum 
4--current density of scalar gravitational waves in the Jordan frame is
\be
T_{0 \mu}=-k_{\mu} ( k_{\nu}\xi^{\nu}) \frac{\varphi}{\phi_0} \; .  
\ee
In the Jordan frame, the energy density and current of spin~0
gravitational waves average to zero on time intervals much longer than the
period of the waves.  However, this is not a solution to the problem,
since one can conceive of scalar gravitational waves with very long
period. For example, gravitational waves from astronomical binary systems
have periods ranging from hours to months (waves from $\mu$--Sco, e.g.,
have period $3 \cdot 10^5$~s). The violation of the weak energy condition
over such macroscopic time scales is unphysical. 

\section{Discussion and conclusions}

The violation of the weak energy condition by scalar--tensor theories
formulated in the Jordan conformal frame makes them unviable descriptions
of classical gravity. Due to the fact that scalar dilatonic fields are
ubiquitous
in superstring and supergravity theories, there is a point in considering
Brans--Dicke theory (and its scalar--tensor generalizations) as toy models
for string theories (e.g. \cite{Cho92,Cho94,Turner93}), and in this case
our
considerations should be reanalyzed, because negative energy states are
not forbidden at the quantum level \cite{Witten82,FordRoman929395}.
However, this context is quite limited, and differs from the usual
classical studies of scalar--tensor theories. 

The reluctance of the gravitational physics community in accepting the
energy argument in favor of the Einstein frame is perhaps due to the fact
that it was formulated in a rather abstract way. The example illustrated
in the present paper shows, in a straightforward way, the
violation of the weak energy condition by wave--like gravitational fields
in Brans--Dicke theory formulated in the Jordan frame, and the viability
of the Einstein frame counterpart of the same theory.  The example is not
academic, since a infrared catastrophe for scalar gravitational waves
would have many observational consequences. One example studied in the
astronomical literature consists of the amplification effect induced by
scalar--tensor gravitational waves on a light beam, which differs in the
Jordan and in the Einstein frame
\cite{Faraoni96}--\cite{BraccoTeyssandier}. 

The argument discussed in this paper for Brans--Dicke theory can be easily
generalized to other scalar--tensor theories. Our conclusions agree with,
and are complementary to, those of Refs.~\cite{Cho92}--\cite{Cho97},
although our approach is different. It has also been pointed out that the
Einstein frame variables $\left( \tilde{g}_{\mu\nu}, \tilde{\phi} \right)
$, but not the Jordan frame variables $\left( g_{\mu\nu}, \phi \right) $,
are appropriate for the formulation of the Cauchy problem
\cite{Cauchyprobl}.

The example presented in this paper agrees with recent studies of the
gravitational collapse to black holes in Brans--Dicke theory
\cite{Scheeletal95}. The noncanonical form of the stress--energy tensor of
the Jordan frame Brans--Dicke scalar is responsible for the violation of
the null energy condition ($R_{\alpha\beta} n^{\alpha} n^{\beta} \geq 0$
for all null vectors $n^{\alpha}$). This causes a {\em decrease} in time
of
the area of the black hole horizon \cite{Scheeletal95}, contrarily to the
behaviour predicted by black hole thermodynamics \cite{Wald84} in general
relativity. In our example, the null energy condition is satisfied, but
there are still pathologies due to the violation of the weak energy
condition. Within the classical context, scalar--tensor theories must be
formulated in the Einstein conformal frame, not in the Jordan one.

\section*{Acknowledgments}

This work was partially supported by EEC grants numbers PSS*~0992 and
CT1*--CT94--0004, and by OLAM, Fondation pour la Recherche Fondamentale,
Brussels.

        \end{document}